\documentclass[final,twocolumn,twoside]{IEEEtran}

\IEEEoverridecommandlockouts
\usepackage{amsmath,mathrsfs}
\usepackage{esvect, amssymb}
\usepackage{epsfig}
\usepackage{epstopdf}
\usepackage{tikz, subcaption, cite}
\usepackage[all]{xy}
\usepackage{xspace,bm}
\usepackage[lined,linesnumbered,algoruled,noend]{algorithm2e}

\newtheorem{theorem}{Theorem}[section]
\newtheorem{definition}[theorem]{Definition}
\newtheorem{assumption}[theorem]{Assumption}
\newtheorem{lemma}[theorem]{Lemma}

\newtheorem{remark}[theorem]{Remark}

\newtheorem{example}[theorem]{Example}
\newtheorem{corollary}[theorem]{Corollary}
\newtheorem{proposition}[theorem]{Proposition}  
\newcommand{\real}{{\mathbb{R}}}

\newcommand{\realnonnegative}{{\mathbb{R}}_{\ge 0}}

\newcommand{\identity}[1]{\mathsf{I}_{#1}}

\newcommand{\ones}{\mathbf{1}}
\newcommand{\zeros}{\mathbf{0}}

\newcommand{\oprocendsymbol}{\hbox{$\bullet$}}
\newcommand{\oprocend}{\relax\ifmmode\else\unskip\hfill\fi\oprocendsymbol}

\newcommand{\map}[3]{#1:#2 \rightarrow #3}

\newcommand{\G}{\mathcal{G}}
\newcommand{\R}{\mathcal{R}}
\newcommand{\E}{\mathcal{E}}

\newcommand{\xnet}{\bm{x}_{\mathcal{R}}}

\newcommand{\xR}{\bm{x}_{\mathcal{R}}}
\newcommand{\dR}{\bm{d}_{\mathcal{R}}}
\newcommand{\AR}{\bar{A}}

\newcommand{\funcset}{\mathfrak{S}}

\renewcommand{\theenumi}{(\roman{enumi})}

\newcommand{\myclearpage}{\clearpage}
 \renewcommand{\myclearpage}{}

 \definecolor{BBlue}{cmyk}{.98,0.10,0,.25}

\begin{document}

\title{Distributed Optimization Under Adversarial Nodes}

\author{Shreyas Sundaram \qquad Bahman Gharesifard\thanks{Shreyas Sundaram (corresponding author) is with the School of Electrical and Computer Engineering at Purdue University, W. Lafayette, IN, 47906, USA. Phone: 1-765-496-0406. Email: \texttt{sundara2@purdue.edu}.  Bahman Gharesifard is with 
the Department of Mathematics and Statistics at Queen's University, Kingston, ON, K7L 3N6, Canada. Phone:  1-613-533-2390. Email: \texttt{bahman@queensu.ca}. 
The work of the second author was partially supported by the Natural Sciences and Engineering Research Council of Canada.
}}

\maketitle

\begin{abstract}
We investigate the vulnerabilities of consensus-based distributed optimization protocols to nodes that deviate from the prescribed update rule (e.g., due to failures or adversarial attacks).  We first characterize certain fundamental limitations on the performance of any distributed optimization algorithm in the presence of adversaries.  We then propose a resilient distributed optimization algorithm that guarantees that the non-adversarial nodes converge to the convex hull of the minimizers of their local functions under certain conditions on the graph topology, regardless of the actions of a certain number of adversarial nodes.  In particular, we provide sufficient conditions on the graph topology to tolerate a bounded number of adversaries in the neighborhood of every non-adversarial node, and necessary and sufficient conditions to tolerate a globally bounded number of adversaries.  For situations where there are up to $F$ adversaries in the neighborhood of every node, we use the concept of maximal $F$-local sets of graphs to provide  lower bounds on the distance-to-optimality of achievable solutions under any algorithm.  We show that finding the size of such sets is NP-hard.
\end{abstract}

\section{Introduction}\label{section:intro}

In recent years, the topic of {\it distributed optimization} has become a canonical problem in the study of networked systems. In this setting, a group of agents equipped with individual objective functions are required to agree on a state that optimizes the sum of these functions. As in the classical consensus problem, the agents can only operate on local information obtained from their neighboring agents, described by a communication network. There is a vast literature devoted to designing distributed algorithms, both in discrete and continuous-time, that guarantee convergence to an optimizer of the sum of the objective functions under reasonable convexity and continuity assumptions~\cite{JNT-DPB-MA:86, MR-RN:04,LX-SB:06, AN-AO:09,PW-MDL:09,AN-AO-PAP:10,BJ-MR-MJ:09,MZ-SM:12, JW-NE:10,JW-NE:11, BG-JC:14-tac, AN-AO:15-tac}.

As outlined above, the predominant assumption in distributed optimization is that all agents cooperate to calculate the global optimizer. In particular, in typical distributed optimization protocols, the individuals update their state via a combination of an agreement term and an appropriately scaled gradient flow of their individual functions.  Given the potential applications of distributed optimization algorithms in large-scale (and safety-critical) cyber-physical systems, and motivated by studies of resilience issues in consensus dynamics (e.g., see \cite{LL-RS-MP:82,NL-96,SS-CN:11-tac,NHV-LT-GL:12,FP-AB-FB:12,HJL-HZ-XK-SS:13}), it is reasonable to ask how vulnerable consensus-based distributed optimization  algorithms are with respect to failure or malicious behavior by certain nodes. In fact, as we argue in this paper, current consensus-based distributed optimization algorithms are easily disrupted by adversarial behavior.  The main objective of this paper is hence to address the issue of resilience of consensus-based distributed optimization dynamics to failure and adversarial behavior, and to refine existing distributed optimization protocols to provide certain safety guarantees against different numbers and types of attackers. The recent work \cite{LS-NV:15} also considers the problem of distributed optimization with adversaries under different assumptions on the graph topology, faulty behavior and classes of functions than the ones that we consider here. The material in this paper substantially extends the conference papers \cite{SS-BG:15-allerton, SS-BG:16-CDC} by providing complete proofs of the results, along with characterizations of the factors that affect the performance of distributed optimization algorithms under adversarial behavior.  The contributions of this paper can be summarized as follows. 

\subsection*{Statement of Contributions}
The first contribution of this paper is to demonstrate fundamental limitations on the performance of {\it any} distributed optimization algorithm in the presence of adversaries. In particular, we show that it is impossible to develop an algorithm that always finds optimal solutions in the absence of adversaries and is at the same time resilient to carefully crafted attacks. 

As our second contribution, we introduce a resilient version of the consensus-based distributed optimization protocol, which we term {\it Local Filtering (LF) Dynamics}, in which the nodes discard the most extreme values in their neighborhood at each time-step. We investigate the capabilities of such protocols under different classes of adversarial behavior, and under the assumption of having an upper bound $ F $ on either the \emph{total} number of adversarial nodes in the network (termed the $F$-total model) or on the \emph{local} number of adversarial nodes in the neighborhood of each non-adversarial node (termed the $F$-local model).  In particular, we provide graph-theoretic sufficient conditions for consensus in scenarios with $ F $-local Byzantine adversaries (which can send different values to different neighbors at each time-step), and necessary and sufficient conditions for scenarios with $ F $-total malicious adversaries (which operate under the wireless broadcast model of communication).   
We utilize two different proof techniques for the two scenarios (each of which provides different insights and capabilities); the first proof relies on properties of products of stochastic matrices for rooted graphs, and relates the consensus value to the limiting left-eigenvector of the subgraph of regular nodes corresponding to eigenvalue $1$.  The second proof relies on characterizing the contracting behavior of the gap between the regular agents with extreme values, and applies even when the graphs are not rooted at each time-step (which can occur under our dynamics, as we demonstrate).   

Our third contribution is to provide a safety guarantee for the proposed LF-dynamics. When the sequence of gradient step-sizes decreases to zero and has infinite $1$-norm (a typical condition in gradient-based optimization dynamics \cite{AN-AO-PAP:10}), we prove that the states of the non-adversarial nodes converge to the convex hull of the minimizers of the individual functions, regardless of the actions taken by the adversarial nodes. 

As our last contribution, we characterize factors that affect the performance of resilient distributed optimization algorithms.  
We provide a bound which shows that for graphs with large so-called \emph{maximum $ F $-local sets}, the performance of resilient algorithms can be poor under the $F$-local adversary model. As a by-product, we prove that the complexity of finding the size of the maximum $ F $-local set is NP-hard. Several examples demonstrate our results. 

\subsection*{Organization} Section~\ref{section:prelim} introduces various mathematical preliminaries. In Section~\ref{sec:dist_opt_consensus}, we review the standard consensus-based distributed optimization algorithm. We describe the adversary model in Section~\ref{section:vulnerabilities}, illustrate vulnerabilities in existing algorithms, and provide fundamental limitations on {\it any} distributed optimization algorithms under such adversarial behavior. We then introduce a class of resilient distributed optimization algorithms in Section~\ref{sec:resilient_consensus_algo}; we provide our main results on consensus under this algorithm in Section~\ref{sec:consensus}, and provide safety guarantees on this algorithm in Section~\ref{sec:safety}.  We identify factors that affect the performance of resilient distributed optimization algorithms  in Section~\ref{sec:performance}, and  conclude in Section~\ref{sec:conclusion}.


\myclearpage 
\section{Mathematical Notation and Terminology}\label{section:prelim}
Let $\real$, $\realnonnegative$,  and $\mathbb{N}$ denote the real,  nonnegative real, and natural numbers, respectively, $ \|\cdot \| $ the Euclidean norm on $ \real^n$, $
\ones =\left[\begin{matrix}1 & 1 & \cdots & 1\end{matrix}\right]'$, $\zeros=\left[\begin{matrix}0 & 0 & \cdots & 0\end{matrix}\right]'$,
and~$ \identity{n} $ the identity matrix in $ \mathbb{R}^{n\times n}$. A matrix $ A \in \real^{n\times n} $ with nonnegative entries is called (row) stochastic if $ A\ones= \ones $. 
Throughout this paper, we are concerned with stochastic matrices whose diagonal entries are bounded away from zero.   For a locally Lipschitz function $\map{f}{\real}{\real}$, we denote the set of subgradients at a given point $x \in \real$ by $\partial f(x)$.

A graph $\G = (V, \E)$ consists of a set of {\it vertices} (or {\it nodes}) $V = \{v_1, v_2, \ldots, v_n\}$, and a set of {\it edges} $\E \subset V \times V$.  The graph is said to be {\it undirected} if $(v_i,v_j) \in \E \Leftrightarrow (v_j,v_i) \in \E$, and {\it directed} otherwise.  The {\it in-neighbors} and {\it out-neighbors} of vertex $v_i \in V$ are denoted by the sets $\mathcal{N}_{i}^{-} \triangleq \{v_j \in V \mid (v_j, v_i) \in \E\}$  and $\mathcal{N}_i^{+} \triangleq \{v_j \in V \mid (v_i,v_j) \in \E\}$, respectively.  The {\it in-degree} and {\it out-degree} of vertex $v_i \in V$ are denoted by $d_i^- \triangleq |\mathcal{N}_i^{-}|$ and $d_i^+ \triangleq |\mathcal{N}_i^{+}|$, respectively.  For undirected graphs, we denote $\mathcal{N}_i = \mathcal{N}_i^- = \mathcal{N}_i^+$ as the {\it neighbors} of vertex $v_i \in V$, and $d_i = d_i^- = d_i^+$ as the {\it degree}.  We denote time-varying graphs, edge sets, and neighbor sets by appending a time-index to those quantities.

A {\it path} from vertex $v_i \in V$ to vertex $v_j \in V$ is a sequence of vertices $v_{k_1}, v_{k_2}, \ldots, v_{k_l}$ such that $v_{k_1} = v_i$, $v_{k_l} = v_j$ and $(v_{k_r}, v_{k_{r+1}}) \in \E$ for $1 \le r \le l-1$.  A graph $ \G=(V,\E) $ is said to be {\it rooted at vertex $ v_i\in V $} if for all vertices $v_j \in V \setminus \{v_i\}$, there a path from $ v_i $ to $v_j$. A graph is said to be {\it rooted} if it is rooted at some vertex $v_i \in V$.    
A graph is {\it strongly connected} if there is a path from every vertex to every other vertex in the graph.  

For any $r \in \mathbb{N}$, a subset $S \subset V$ of vertices is said to be {\it $r$-local} if $|\mathcal{N}_i^- \cap S| \le r$ for all $v_i \in V \setminus S$.  In other words, if $S$ is $r$-local, there are at most $r$ vertices from $S$ in the in-neighborhood of any vertex from $V \setminus S$. A {\it maximum $r$-local} set is an $r$-local set of largest cardinality (i.e., there are no $r$-local sets of larger size).  A subset $S \subset V$ of vertices is said to be {\it $r$-reachable} if there exists a vertex $v_i \in S$ such that $|\mathcal{N}_i^{-}\setminus S| \ge r$.  In other words, $S$ is $r$-reachable if it contains a vertex that has at least $r$ in-neighbors from outside $S$.  

The following definitions of robust graphs will play a role in our analysis.

\begin{definition}[$r$-robust graphs]
For $r \in \mathbb{N}$, graph $\G$ is said to be {\it $r$-robust} if for all pairs of disjoint nonempty subsets $S_1, S_2 \subset V$, at least one of $S_1$ or $S_2$ is $r$-reachable.
\end{definition}

\begin{definition}[$(r,s)$-robust graphs]
For $r, s \in \mathbb{N}$, a graph is said to be $(r,s)$-robust if for all pairs of disjoint nonempty subsets $S_1, S_2 \subset V$, at least one of the following conditions holds:
\begin{enumerate}
\item All nodes in $S_1$ have at least $r$ neighbors outside $S_1$.
\item All nodes in $S_2$ have at least $r$ neighbors outside $S_2$.
\item There are at least $s$ nodes in $S_1 \cup S_2$ that each have at least $r$ neighbors outside their respective sets.
\end{enumerate}
\label{def:rs_robust}
\end{definition}

The above definitions capture the idea that given any two disjoint nonempty  subsets of nodes in the network, there are a certain number of nodes within those sets that each have a sufficient number of neighbors outside their respective sets.  This notion will play a key role in the resilient dynamics that we propose in this paper, where nodes choose to discard a certain number of their neighbors in order to mitigate adversarial behavior.  Note that $(r,1)$-robustness is equivalent to $r$-robustness.  The following result (from Lemma 6 and Lemma 7 in \cite{HJL-HZ-XK-SS:13}) will be useful for our analysis.

\begin{lemma}\label{lem:r_robust_rooted}
Suppose a graph $\G$ is $r$-robust.  Let $\G'$ be a graph obtained by removing $r-1$ or fewer incoming edges from each node in $\G$.  Then $\G'$ is rooted.
\end{lemma}

Further details on the above notions of robustness can be found in \cite{HJL-HZ-XK-SS:13, HZ-EF-SS:15}.

\myclearpage
\section{Review of Consensus-Based Distributed Optimization}
\label{sec:dist_opt_consensus}

Consider a network consisting of $n$ agents $V=\{v_1,\dots, v_n\} $ whose communication topology is a potentially time-varying graph~$\G(t)=(V,\E(t)) $. An edge
$(v_i,v_j) \in \E(t) $ indicates that $v_j$ can receive information
from $v_i$ at time-step $t \in \mathbb{N}$.  For each $i \in \{1,\ldots, n\} $, let $
\map{f_i}{\real}{\real} $ be locally Lipschitz and convex,
and only available to agent $v_i$. The objective is for the agents to solve,
in a distributed way (i.e., by exchanging information only with their immediate neighbors), the global optimization problem\footnote{In order to tackle the complexities associated with adversarial behavior, we restrict attention to scalar unconstrained optimization problems throughout the paper.} 
\begin{align}\label{eq:dis_opt}
  \mathrm{minimize} \quad f(x)=\frac{1}{n}\sum_{i=1}^n f_i(x) .
\end{align}

A common approach to solve this problem is to use a synchronous iterative consensus-based protocol in which agents use a combination of consensus dynamics and gradient flow to find a minimizer of $ f $ \cite{AN-AO:09,AN-AO-PAP:10,AN-AO:10-tutorial}.   Specifically, at each time-step $t \in \mathbb{N}$, each agent $v_i \in V$ has an estimate $x_i(t) \in \mathbb{R}$ of the solution to the problem \eqref{eq:dis_opt}.  Each agent $v_i \in V$ sends its estimate to its out-neighbors, receives the estimates of its in-neighbors, and updates its estimate as \cite{AN-AO:09}
\begin{equation}
x_i(t+1) = a_{ii}(t)x_i(t) + \sum_{v_j \in \mathcal{N}_i^-(t)}a_{ij}(t)x_j(t) - \alpha_td_i(t).
\label{eq:dist_opt_update}
\end{equation}
In the above update rule, $a_{ij}(t)$, $v_j \in  \{v_i\} \cup \mathcal{N}_i^-(t)$, are a set of nonnegative real numbers satisfying $a_{ii}(t) + \sum_{v_j \in   \mathcal{N}_i^-(t)}a_{ij}(t)  = 1$.  In other words, the first portion of the right hand side is a {\it consensus step}, representing a weighted average of the estimates in node $v_i$'s neighborhood.  The quantity $d_i(t)$ is a subgradient of $f_i$, evaluated at $a_{ii}(t)x_i(t) + \sum_{v_j \in \mathcal{N}_i^-(t)}a_{ij}(t)x_j(t)$.  Finally, $\{\alpha_t\}_{t \in \mathbb{N}}$, is the {\it step-size} sequence corresponding to the influence of the subgradient on the update rule at each time-step.  In this sense, the last term in the above expression represents a {\it gradient step}.  

The dynamics \eqref{eq:dist_opt_update} can be represented compactly as follows.  Let 
\begin{align*}
x(t) &\triangleq \left[\begin{matrix}x_{1}(t) & x_2(t) & \cdots & x_{n}(t)\end{matrix}\right]' \in \real^{n}, \\
d(t) &\triangleq \left[\begin{matrix}d_{1}(t) & d_2(t) & \cdots & d_{n}(t)\end{matrix}\right]' \in \real^{n} 
\end{align*}
be the vector of states and subgradients of the nodes at time-step $t$, respectively.  Let $A(t) \in \real_{\ge 0}^{n \times n}$ be the matrix such that for each $(v_j,v_i) \in \E(t)$, the $ (i,j) $-th entry of $A(t)$ is ${a}_{ij}(t)$ given in~\eqref{eq:dist_opt_update}, the diagonal elements of $A(t)$ are the self-weights $a_{ii}(t)$, and all other entries are set to zero. Then \eqref{eq:LF_dynamics_regular} can be written as
\begin{equation}\label{eq:dist_opt_matrix_form}
x(t+1) = A(t)x(t) - \alpha_td(t),
\end{equation}
for $t \in \mathbb{N}$.  Note that each row of $A(t)$ sums to $1$ at each time-step, and thus $A(t)$ is row-stochastic.  It is easy to observe that 
\begin{align}
x(t+1)&=A(t)A(t-1) \cdots A(0) x(0)\label{eq:matrix-form-expanded}\\
&-\sum_{s=1}^{t} A(t) A(t-1) \cdots A(s)\alpha_{s-1}d(s-1) - \alpha_td(t). \nonumber
\end{align}
For notational convenience, we define $\Phi(t,s) \triangleq A(t)A(t-1) \cdots A(s)$ for $t \ge s$, and $\Phi(t,s) \triangleq 0$ for $t < s$.  Thus, \eqref{eq:matrix-form-expanded} becomes
\begin{equation*}
x(t+1) = \Phi(t,0)x(0) - \sum_{s=1}^{t}\Phi(t,s)\alpha_{s-1}d(s-1) - \alpha_td(t).
\end{equation*}
There are some commonly-used assumptions that are made on the weights in~\eqref{eq:dist_opt_update}, which we encapsulate below.

\begin{assumption}[Lower Bounded Weights]
There exists a constant $\eta > 0$ such that for all $ t \in \mathbb{N}$ and $v_i \in V$,  if $v_j \in \{v_i\} \cup \mathcal{N}_i^-(t)$, then $a_{ij}(t) \ge \eta$.
\label{as:bounded_weights}
\end{assumption}

\begin{assumption}[Double Stochasticity]
For all $t \in \mathbb{N}$ and $v_i \in V$, the weights satisfy $a_{ii}(t) + \sum_{v_j \in  \mathcal{N}_i^+(t)}a_{ji}(t)  = 1$.  
\label{as:double_stochastic}
\end{assumption}

The following result is a special case of the results of \cite{AN-AO-PAP:10} for graphs that are strongly connected at each time-step.

\begin{proposition}\label{prop:Nedic10_opt}
Suppose the network $\G(t)$ is strongly connected at each time-step.  Suppose the subgradients of each of the local functions $f_i$ are bounded, i.e., there exists $ L \in \mathbb{R}_{> 0}$ such that $\|d\| \le L$, for all $d \in \partial f_i(x)$ and $x \in \mathbb{R}$.  Consider the update rule \eqref{eq:dist_opt_update}, and suppose the weights satisfy Assumption~\ref{as:bounded_weights} and Assumption~\ref{as:double_stochastic}.  Let the step-sizes satisfy $\sum_{t \in \mathbb{N}} \alpha_t = \infty$ and $\sum_{t \in \mathbb{N}}\alpha_t^2 < \infty$.  Then there is a minimizer $x^{\ast}$ of \eqref{eq:dis_opt} such that
$$
\lim_{t\rightarrow\infty}\|x_i(t) - x^{\ast}\| = 0,
$$
for all $v_i \in V$.
\end{proposition}

The above result shows that the update rule \eqref{eq:dist_opt_update} allows the nodes in the network to distributively solve the global optimization problem \eqref{eq:dis_opt}.
Our main objective in this paper is to investigate the vulnerabilities of such protocols to nodes that {\it deviate} from the prescribed update rule (e.g., due to failures or adversarial attacks), and to develop a resilient distributed optimization algorithm that has provable safety guarantees in the presence of such deviations.   To do this, it will be helpful to first  generalize the above analysis to handle cases where the weights are not doubly-stochastic.  

\subsection{Scenarios with Non-Doubly-Stochastic Weights}
\label{sec: general_opt}

Here, we will establish convergence of the node states under the dynamics \eqref{eq:dist_opt_update} under certain classes of non-doubly-stochastic consensus weights.  
At each time-step $t \in \mathbb{N}$, let $A(t) \in \mathbb{R}_{\ge 0}^{n\times n}$ be the matrix containing the weights $a_{ij}(t)$.  Note that $a_{ij}(t) = 0$ if $v_j \notin \{v_i\}\cup \mathcal{N}_i^-(t)$.  Suppose there exists some constant $\beta > 0$ such that at each time-step $t \in \mathbb{N}$, $A(t)$ has a rooted subgraph that has edge-weights lower-bounded by $\beta$, and diagonal elements lower-bounded by $\beta$.  Let $\Phi(t,s) \triangleq A(t)A(t-1)\cdots A(s)$ for $t \ge s \ge 0$. Using the fact that $A(t)$ has a rooted subgraph, and with an argument similar to the one in~\cite{MC-ASM-BDOA:08} which we omit  here, for each $s \in \mathbb{N}$, there exists a stochastic vector $\mathbf{q}_s$ such that 
\begin{equation}
\lim_{t\rightarrow\infty} \Phi(t,s) = \ones{\mathbf{q}'_s}.
\label{eq:q_s_def}
\end{equation}
Noting that $\Phi(t,s)  = \Phi(t,s+1)A(s)$, we have that
\begin{equation}
\mathbf{q}'_s = \mathbf{q}'_{s+1}A(s), 
\label{eq:q_s_recursion}
\end{equation}
for all $s \in \mathbb{N}$.

For each $t \in \mathbb{N}$, let $x(t) \in \mathbb{R}^n$ be the state vector for the network, and define the quantity 
\begin{equation}
y(t) \triangleq \mathbf{q}_t'x(t)
\label{eq:ydef}
\end{equation}
 (i.e., $y(t)$ is a convex combination of the states of the nodes at time-step $t$).  Using the above definition, we have the following convergence result.  The proof of this result closely follows the proof for doubly-stochastic weights provided in \cite{AN-AO-PAP:10}, with the main difference being in the use of the vector $\mathbf{q}_t$ at  appropriate points.

\begin{lemma}\label{lem:consensus_to_y}
 Consider the network $\G(t) = (V,\E(t))$. Suppose that the functions $f_i$, $v_i \in V$, have subgradients bounded by some constant $L$, and that the nodes run the dynamics \eqref{eq:dist_opt_update}.  Assume that there exists a constant $\beta > 0$ such that at each time-step $t \in \mathbb{N}$, the weight matrix $A(t)$ has diagonal elements lower bounded by $\beta$ and contains a rooted subgraph whose edge weights are lower bounded by $\beta$.   Let $y(t)$ be the corresponding sequence defined in \eqref{eq:ydef}.
\begin{enumerate}
\item If $\alpha_t \rightarrow 0$ as $t \rightarrow \infty$, then
$$
\limsup_{t \rightarrow\infty}\left\| x(t)- \ones y(t)\right\| =  0.
$$
\item If $\sum_{t = 1}^{\infty}\alpha_t^2 < \infty$, then
$$
\sum_{t=1}^{\infty}\alpha_t\|x(t)-\ones y(t)\| < \infty.
$$
\item  If each matrix $A(t)$, $t\in\mathbb{N}$ has a common left-eigenvector $\mathbf{q}'$ corresponding to eigenvalue $1$, and the step-sizes satisfy $\sum\alpha_t = \infty$ and $\sum \alpha_t^2 < \infty$,  then
$$
\lim_{t\rightarrow\infty}\|x_i(t)-x^*\| = 0
$$
for all $v_i \in V$, where $x^*$ is a minimizer of $\sum_{i = 1}^nq_if_i$, with  $q_i$ being the $i$-th entry of $\mathbf{q}'$. 
\end{enumerate}
\end{lemma}

Note that if the matrices $A(t)$ do not have a common left-eigenvector, convergence to a constant value is not guaranteed under the dynamics \eqref{eq:dist_opt_update} (unlike in standard consensus dynamics without the gradient terms).  To see this, consider two row-stochastic matrices $A_1$ and $A_2$, each with rooted subgraphs and nonzero diagonal elements, with different left eigenvectors $\mathbf{q}_1'$ and  $\mathbf{q}_2'$, respectively, for eigenvalue $1$.  Select the functions for the nodes such that $\sum q_{1i}f_i$ and $\sum q_{2i}f_i$ have different minimizers, where $q_{ij}$ is the $j$-th component of $\mathbf{q}_i$.  Then, if the dynamics evolve according to matrix $A_1$ for a sufficiently large period of time, all nodes will approach the minimizer of $\sum q_{1i}f_i$, regardless of the initial conditions.  Similarly, if the dynamics evolve according to the matrix $A_2$ for a sufficiently large period of time, all nodes will approach the minimizer of $\sum q_{2i}f_i$, again regardless of the initial conditions.  Thus, by appropriately switching between the matrices $A_1$ and $A_2$, the nodes will oscillate between the two different minimizers.

With these results on distributed optimization in hand, we now turn our attention to the effect of adversaries on the optimization dynamics.  


\myclearpage
\section{Adversary Model and Vulnerabilities of Distributed Optimization Algorithms}\label{section:vulnerabilities}

Henceforth, we will assume that the underlying graph $\G$ is time-invariant in order to focus on issues pertaining to resilience to adversarial behavior.  However, as we will see later, our proposed algorithm will utilize time-varying (and state-dependent) weights which can be viewed as inducing time-varying subgraphs of the underlying graph $\G$.

\subsection{Adversary Model}
We partition the set of nodes $V$ into two subsets:   a set of {\it adversarial nodes} $\mathcal{A}$, and a set of {\it regular nodes} $\mathcal{R} = V \setminus \mathcal{A}$.  The system undergoes the following sequence of steps:

 \begin{enumerate}
\item Each node $v_i \in V$ draws a private function $f_i$ that is locally Lipschitz and convex.
\item A set of nodes $\mathcal{A}\subset V$ is selected by an attacker to be adversarial.  The attacker is allowed to know the entire network topology and the private functions assigned to all of the nodes when selecting the set $\mathcal{A}$.  
\item The regular nodes commence running the distributed optimization algorithm.
\end{enumerate}

The regular nodes will exactly follow any algorithm that is prescribed.  The adversarial nodes, on the other hand, can update their states in a completely arbitrary (potentially worst-case and coordinated) manner.   We will classify adversaries in terms of their number, locations, and types of misbehavior, as follows.

\begin{definition}[$F$-total vs. $F$-local]
For $F \in \mathbb{N}$, we say that the set of adversaries $\mathcal{A}$ is an $F$-total set if $|\mathcal{A}| \le F$, and an $F$-local set if $|\mathcal{N}_i^- \cap \mathcal{A}| \le F$, for all $v_i \in \mathcal{R}$.
\end{definition}

\begin{definition}[Malicious vs. Byzantine]
We say that an adversarial node is \emph{malicious} if it sends the same value to all of its out-neighbors at each time-step (i.e., it follows the wireless broadcast model of communication).  We say that an adversarial node is \emph{Byzantine} if it is capable of sending different values to different neighbors at each time-step (i.e., it follows the wired point-to-point model of communication).      
\end{definition}

Note that malicious adversaries are a special case of Byzantine adversaries, and similarly, $F$-total adversaries are a special case of $F$-local adversaries.   
 
  
\subsection{Attacking Consensus-Based Distributed Optimization Algorithms}
 
We start with the following result showing that it is extremely simple for even a single adversarial node (either malicious or Byzantine) to disrupt dynamics of the form \eqref{eq:dist_opt_update}.  
 
\begin{proposition}
 Consider the network $\G = (V,\E)$, and let there be a single adversarial node $\mathcal{A} = \{v_n\}$.  Suppose the network is rooted at $v_n$.  Then if $v_n$ keeps its value fixed at some constant $\bar{x} \in \mathbb{R}$ and the step-sizes satisfy $\alpha_t \rightarrow 0$, all regular nodes will asymptotically converge to $\bar{x}$ when following the distributed optimization dynamics \eqref{eq:dist_opt_update}.
\end{proposition}

\begin{IEEEproof}
 Since the adversarial node keeps its value fixed for all time, its update can be modeled as
 $$
 x_n(t+1) = x_n(t)
 $$
 for all $t \in \mathbb{N}$, with $x_n(0) = \bar{x}$.  Thus, the global distributed optimization dynamics take the form shown in \eqref{eq:dist_opt_matrix_form}, with 
 $$
 A(t) = \left[\begin{matrix}A_{\mathcal{R},\mathcal{R}}(t) & A_{\mathcal{R},\mathcal{A}}(t)\\ 0 & 1\end{matrix}\right],
 $$
 where $A_{\mathcal{R},\mathcal{R}}(t)$ is the matrix containing the weights placed by regular nodes on other regular nodes during the update \eqref{eq:dist_opt_update}, and $A_{\mathcal{R},\mathcal{A}}(t)$ is a vector containing the weights placed by regular nodes on the adversarial node's value.  Since (i) the graph contains a spanning tree rooted at $v_n$, (ii) all weights used by the regular nodes on their neighbors (and own values) are bounded away from zero, and (iii) all matrices $A(t)$ have a common left-eigenvector $\mathbf{q}' = \left[\begin{matrix} 0_{1 \times {n-1}} & 1\end{matrix}\right]$, the first part of Lemma~\ref{lem:consensus_to_y} indicates that all regular nodes will converge to $y(t) = \mathbf{q}'x(t) = x_n(t) =\bar{x}$.  
\end{IEEEproof}

The above phenomenon is entirely analogous to the behavior that occurs under ``stubborn'' agents in standard consensus dynamics (e.g.,  \cite{JG-RS:14,EY-AO-DA-AS-AS:13}).  


\subsection{Fundamental Limitations on Any Resilient Distributed Optimization Algorithm}

The previous result shows that consensus-based distributed optimization algorithms can be co-opted by an adversary simply fixing its value at some constant.  It is plausible that  this type of simple misbehavior can be detected via an appropriate mechanism.    However, it is easy to argue as follows that under mild conditions on the class of objective functions at each node, an adversary can {\it always} behave in way as to avoid detection, while arbitrarily affecting the outcome of the distributed optimization.

\begin{theorem}
  Suppose the local objective functions at each node are convex with bounded subgradients, but otherwise completely arbitrary.  Suppose $\Gamma$ is a distributed algorithm that guarantees that all nodes calculate the global optimizer of problem \eqref{eq:dis_opt} when there are no adversarial nodes.  Then a single adversary can cause all nodes to converge to any arbitrary value when they run algorithm $\Gamma$, and furthermore, will remain undetected.
  \label{thm:fund}
\end{theorem}

\begin{IEEEproof}
 Without loss of generality, let $v_n$ be an adversarial node.  Let each node  $ v_i \in V$ have local function $f_i $.  Suppose node $v_n$ wishes all nodes to calculate some value $\bar{x}$ as an outcome of running the algorithm $\Gamma$.  Node $v_n$  chooses a convex function $\bar{f}_n$ such that $\partial \bar{f}_n(\bar{x}) = - \sum_{v_i \in V \setminus \{v_n\}}\partial f_i(\bar{x})$, with gradient capped at a sufficiently large value.  Thus, the global minimizer of the function $\frac{1}{n}\left(\sum_{v_i \in V\setminus\{v_n\}}f_i + \bar{f}_n\right)$ is $\bar{x}$.  Now,  node $v_n$ participates in algorithm $\Gamma$ by pretending its local function is $\bar{f}_n$ instead of $f_n$.   Since $\bar{f}_n$ is a legitimate function that could have been assigned to $v_n$, this scenario is indistinguishable from the case where $v_n$ is a regular node, and thus this misbehavior cannot be detected.  Thus, algorithm $\Gamma$ must cause all nodes to calculate $\bar{x}$ under this misbehavior.  
\end{IEEEproof} 
      
The above theorem applies to {\it any} algorithm that is guaranteed to output the globally optimum value in the absence of adversaries.  The takeaway point is that there is a tradeoff between optimality and resilience: any algorithm that always finds optimal solutions in the absence of adversaries (under mild assumptions on the class of local functions) can also be arbitrarily co-opted by an adversary.

In the next section, we build on the insights gained from the above characterizations of fundamental limitations, and propose a modification of the standard consensus-based distributed optimization algorithm that provides certain {\it safety} guarantees in the face of arbitrary adversarial behavior.

\myclearpage
\section{A Resilient Consensus-Based Distributed Optimization Protocol}
\label{sec:resilient_consensus_algo}

Suppose that the adversarial nodes are restricted to form an $F$-local set, where $F$ is a nonnegative integer.  The regular nodes do not know which (if any) of their neighbors are adversarial.  Suppose that at each time-step $t \in \mathbb{N}$, each regular node $v_i \in \mathcal{R}$ performs the following actions in parallel with the other regular nodes:
\begin{enumerate}
\item Node $v_i$ gathers the states $\{x_j(t), v_j \in \mathcal{N}_i^-\}$ of its in-neighbors.
\item Node $v_i$ sorts the gathered values and removes the $F$ highest and $F$ smallest values that are larger and smaller than its own value, respectively.  If there are fewer than $F$ values higher (resp. lower) than its own value, $v_i$ removes all of those values.  Ties in values are broken arbitrarily.  Let $\mathcal{J}_i(t) \subset \mathcal{N}_i^{-}$ be the set of in-neighbors of $v_i$ whose states were retained by $v_i$ at time-step $t$.  
\item Node $v_i$ updates its state as
\begin{equation}\label{eq:LF_dynamics}
x_i(t+1)= a_{ii}(t)x_i(t) + \sum_{v_j \in \mathcal{J}_i(t)} a_{ij}(t) x_j(t) - \alpha_td_i(t),
\end{equation}
where $d_i(t)$ is a subgradient of $f_i$ evaluated at $a_{ii}x_i(t) + \sum_{v_j \in \mathcal{J}_i(t)} a_{ij}(t) x_j(t)$, and $\{\alpha_t\}_{t \in \mathbb{N}}$ is a nonnegative step-size sequence.  
At each time-step $t$ and for each $v_i \in \mathcal{R}$, the weights  $a_{ij}(t)$, $v_j \in \{v_i\} \cup \mathcal{J}_i(t)$, are lower-bounded by some strictly positive real number $\eta$ and sum to $1$ (i.e., they specify a convex combination). 
\end{enumerate}

The adversarial nodes are allowed to update their states however they wish.   Note that the above dynamics are {\it purely-local} in the sense that they do not require the regular nodes to know anything about the network topology (other than their own in-neighbors).   Also note that even when the underlying network $\G$ is time-invariant, the filtering operation induces {\it state-dependent switching} (i.e., the effective in-neighbor set $\mathcal{J}_i(t)$ is a function of the states of the in-neighbors of $v_i$ at time-step $t$).  In case a regular node $v_i$ has a Byzantine neighbor $v_j$, we abuse notation and take the value $x_j(t)$ in the update equation \eqref{eq:LF_dynamics} to be the value received from node $v_j$ (i.e., it does not have to represent the true state of node $v_j$).

We will refer to the above dynamics as {\it Local Filtering (LF) Dynamics} with parameter $F$.   Local filtering operations of the above form have been previously studied in the context of resilient consensus dynamics (i.e., outside of distributed optimization) in \cite{HJL-HZ-XK-SS:13,NHV-LT-GL:12,DD-ANL-SSP-EWS-WEW:86}.  However, the presence of the gradient terms in the dynamics \eqref{eq:LF_dynamics} adds additional complexity that precludes the proof techniques from \cite{HJL-HZ-XK-SS:13} from being directly applied, and thus we will analyze these dynamics in the remainder of the paper, and show that they are resilient to adversarial behavior under certain conditions on the network topology.  

\subsection{A Mathematically Equivalent Representation of Local Filtering Dynamics}

Since we are concerned with understanding the evolution of the states of the regular nodes in our analysis, it will be useful to consider a {\it mathematically equivalent} representation of the dynamics \eqref{eq:LF_dynamics} that only involves the states of the regular nodes.  The key idea of the proof of the following proposition is from \cite{NHV:12-arxiv}, which considered a slightly different version of the local filtering dynamics in the context of distributed consensus.  Here, we provide a somewhat simpler proof, adapted for the version of the dynamics that we are considering.

 \begin{proposition}\label{prop:LF_dynamics_regular}
 Consider the network $\G = (V,\E)$, with a set of regular nodes $\mathcal{R}$ and a set of adversarial nodes $\mathcal{A}$.  Suppose that $\mathcal{A}$ is an $F$-local set, and that each regular node has at least $2F+1$ in-neighbors.  Then the update rule \eqref{eq:LF_dynamics} for each node $v_i \in \mathcal{R}$ is mathematically equivalent to 
 \begin{equation}\label{eq:LF_dynamics_regular}
 x_i(t+1)= \bar{a}_{ii}(t)x_i(t) + \sum_{v_j \in  \mathcal{N}_i^{-}\cap\mathcal{R}} \bar{a}_{ij}(t) x_j(t) - \alpha_td_i(t),
 \end{equation} 
 where the nonnegative weights $\bar{a}_{ij}(t)$ satisfy the following properties at each time-step $t$:  
 \begin{enumerate}
 \item $\bar{a}_{ii}(t) + \sum_{v_j \in \mathcal{N}_i^{-}\cap\mathcal{R}} \bar{a}_{ij}(t) = 1$.
 \item $\bar{a}_{ii}(t) \ge \eta$ and at least $|\mathcal{N}_{i}^-| - 2F$ of the other weights are lower bounded by $\frac{\eta}{2}$.
 \end{enumerate}
 \end{proposition}

 \begin{IEEEproof} 
Consider a regular node $v_i \in \mathcal{R}$.  We will prove the result by providing a procedure to construct the weights $\bar{a}_{ij}(t)$ described in the statement, starting from the weights $a_{ij}(t)$ in the LF dynamics \eqref{eq:LF_dynamics}.  To facilitate this, we define two different partitions of the in-neighbors of $v_i$.  For the first partition, define the sets $\mathcal{U}_i(t)$, $\mathcal{J}_i(t)$ and $\mathcal{L}_i(t)$, where $\mathcal{U}_i(t)$ (resp. $\mathcal{L}_i(t)$) contains the nodes with the highest (resp. lowest) values that were removed by node $v_i$ after the filtering operation.    For the second partition, define the sets $\bar{\mathcal{U}}_i(t)$, $\bar{\mathcal{J}}_i(t)$ and $\bar{\mathcal{L}}_i(t)$, where $\bar{\mathcal{U}}_i(t)$ and $\bar{\mathcal{L}}_i(t)$ contain the highest and lowest $F$ values in node $v_i$'s neighborhood at time-step $t$, respectively.  The set $\bar{\mathcal{J}}_i(t)$ contains the remaining values.  Thus, we have $\mathcal{U}_i(t) \subseteq \bar{\mathcal{U}}_i(t)$, $\bar{\mathcal{J}}_i(t) \subseteq \mathcal{J}_i(t)$, and $\mathcal{L}_i(t) \subseteq \bar{\mathcal{L}}_i(t)$.

Define $\bar{a}_{ii}(t) = a_{ii}(t)$ and $\bar{a}_{ij}(t) = a_{ij}(t)$ for $v_j \in \mathcal{J}_i(t) \cap \mathcal{R}$.  Set $\bar{a}_{ij}(t) = 0$ for $v_j \in \mathcal{R} \setminus \mathcal{J}_{i}(t)$.  

If there are no adversarial nodes in $\mathcal{J}_i(t)$ (i.e., $\mathcal{J}_i(t) = \mathcal{J}_i(t) \cap \mathcal{R}$), then the construction of the weights $\bar{a}_{ij}(t)$ for node $v_i$ is complete.  Specifically, we have
$$
\bar{a}_{ii}(t) + \sum_{v_j \in\mathcal{N}_{i}^- \cap \mathcal{R}} \bar{a}_{ij}(t) = {a}_{ii}(t) + \sum_{v_j \in\mathcal{J}_{i}(t)} {a}_{ij}(t) = 1,
$$ 
which satisfies the first condition in the proposition.  Furthermore, since $|\mathcal{J}_i(t)| \ge |\mathcal{N}_i^-| - 2F$ and each of the weights are lower bounded by $\eta$, this satisfies the second condition in the proposition.

Now consider the case where there are one or more adversarial nodes  in $\mathcal{J}_i(t)$.  We consider adversarial nodes in $\mathcal{J}_i(t)\setminus\bar{\mathcal{J}}_i(t)$ and $\bar{\mathcal{J}}_i(t)$ separately.  

Consider any adversarial node $v_m \in \mathcal{J}_i(t)\setminus\bar{\mathcal{J}}_i(t)$, and let $x_m(t)$ be the value received by node $v_i$ from $v_m$.   Since $v_i$ did not discard $v_m$'s value, it must be the case that there are either $F$ values that are higher than $x_m(t)$ in $v_i$'s neighborhood, or $v_i$'s own value is higher than $x_m(t)$.  Similarly, there must either be $F$ values that are lower than $x_m(t)$ in $v_i$'s neighborhood, or $v_i$'s own value is lower than $x_m(t)$.  Since there are at most $F$ adversarial nodes in $v_i$'s neighborhood, we see that there is a pair of regular nodes $v_u, v_l \in \mathcal{N}_i^-\cup\{v_i\}$ with $x_l(t) \le x_m(t) \le x_u(t)$.  Thus, the term $a_{im}(t)x_m(t)$ in \eqref{eq:LF_dynamics} can be written as 
$$
a_{im}(t)x_m(t) = a_{im}(t)\gamma_mx_u(t) + a_{im}(t)(1-\gamma_m)x_l(t)
$$
for some $\gamma_m \in [0,1]$.  By updating the weights $\bar{a}_{iu}(t)$ and $\bar{a}_{il}(t)$ as $\bar{a}_{iu}(t) \leftarrow \bar{a}_{iu}(t) + a_{im}(t)\gamma_m$ and $\bar{a}_{il}(t) \leftarrow \bar{a}_{il}(t) + a_{im}(t)(1-\gamma_m)$, respectively, the contribution of the adversarial node $v_m \in \mathcal{J}_i(t)\setminus\bar{\mathcal{J}}_i(t)$ in \eqref{eq:LF_dynamics} is transformed into contributions by two regular nodes.  We do this for each adversarial node in $ \mathcal{J}_i(t)\setminus\bar{\mathcal{J}}_i(t)$.

Now consider the set $\bar{\mathcal{J}}_i(t)$, containing $|\mathcal{N}_i^- - 2F|$ nodes.  If there are no adversarial nodes in  $\bar{\mathcal{J}}_i(t)$, then the construction of the weights $\bar{a}_{ij}(t)$ is complete and both conditions in the proposition are satisfied (since the weights assigned to the regular nodes in $\bar{\mathcal{J}}_i(t)$ satisfy the second condition in the proposition by each being larger than $\eta$).  
 
Thus suppose that there are $K$ adversarial nodes in the set $\bar{\mathcal{J}}_i(t)$, where $1 \le K \le F$ (recall that the set of adversarial nodes is assumed to be $F$-local).  Then there must be at least $K$ regular nodes in the set $\bar{\mathcal{U}}_i(t)$, and at least $K$ regular nodes in the set $\bar{\mathcal{L}}_i(t)$.  Label the $K$ adversarial nodes in $\bar{\mathcal{J}}_i(t)$ as $\{v_{m_1}, v_{m_2}, \ldots, v_{m_K}\}$, with corresponding states $x_{m_1}(t), x_{m_2}(t), \ldots, x_{m_K}(t)$.  Pick any $K$ regular nodes in $\bar{\mathcal{U}}_i(t)$ and any $K$ regular nodes in $\bar{\mathcal{L}}_i(t)$, and label them as $\{v_{u_1}, v_{u_2}, \ldots, v_{u_K}\}$, and  $\{v_{l_1}, v_{l_2}, \ldots, v_{l_K}\}$, respectively.  We will label the states of these nodes as $x_{u_1}(t), x_{u_2}(t), \ldots, x_{u_K}(t)$, and $x_{l_1}(t), x_{l_2}(t), \ldots, x_{l_K}(t)$, respectively.    By definition, we have $x_{l_j}(t) \le x_{m_j}(t) \le x_{u_j}(t)$ for all $1 \le j \le K$.   Thus for each $j \in \{1, 2, \ldots, K\}$, we can write
$$
x_{m_j}(t) = \gamma_{j}x_{l_j}(t) + (1-\gamma_j)x_{u_j}(t),
$$
where $0 \le \gamma_j \le 1$.  In other words, the state of the adversarial node $v_{m_j}$ is a convex combination of the states of the regular nodes $v_{u_j}$ and $v_{l_j}$.  Note that either $\gamma_j$ or $(1-\gamma_j)$ must be at least equal to $0.5$.  

As before, update the weights $\bar{a}_{il_j}(t)$ and $\bar{a}_{iu_j}(t)$ as $\bar{a}_{il_j}(t) \leftarrow \bar{a}_{il_j}(t) + a_{im_j}(t)\gamma_j$ and $\bar{a}_{iu_j}(t) \leftarrow \bar{a}_{iu_j}(t) +  a_{im_j}(t)(1-\gamma_j)$ for $j \in \{1, 2, \ldots, K\}$.  In other words, we split the value of the weight that was assigned to the adversarial node $m_j$ among the regular nodes $l_j$ and $u_j$, according to the proportions $\gamma_j$ and $(1-\gamma_j)$.  Note that at least $K$ of the nodes in $\{v_{u_1}, v_{u_2}, \ldots, v_{u_K}\} \cup \{v_{l_1}, v_{l_2}, \ldots, v_{l_K}\}$ get assigned a weight that is lower bounded by $\frac{\eta}{2}$ (since either $\gamma_j$ or $(1-\gamma_j)$ is at least $0.5$).  Since the weight associated to each adversarial node is split according to a convex combination to a pair of regular nodes in $\mathcal{N}_{i}^{-}\setminus\bar{\mathcal{J}}_i(t)$, we see that the first condition in the proposition is satisfied.  Finally, since  $\bar{a}_{ij}(t) \ge a_{ij}(t) \ge \eta$ for $v_j \in \bar{\mathcal{J}}_i(t)\cap\mathcal{R}$, this ensures that $|\bar{\mathcal{J}}_i(t)|-K = |\mathcal{N}_i^-|-2F-K$ weights are lower bounded by $\eta$.  As discussed above, the splitting of the adversarial nodes' weights ensures that an additional $K$ regular nodes are assigned a weight that is lower bounded by $\frac{\eta}{2}$.  Thus, in total, there are at least $|\mathcal{N}_i^-|-2F$ weights (other than $\bar{a}_{ii}(t)$) that are lower bounded by $\frac{\eta}{2}$, concluding the proof.
\end{IEEEproof}

 
We emphasize again that the regular nodes run the dynamics \eqref{eq:LF_dynamics} (which does not require them to know which of their neighbors is adversarial); the dynamics \eqref{eq:LF_dynamics_regular} are {\it mathematically equivalent} to the dynamics \eqref{eq:LF_dynamics} due to the nature of the local filtering that is done by each regular node, and will lead to certain insights that we will leverage.

Henceforth, we assume without loss of generality that the regular nodes are arranged first in the ordering of the nodes, and define 
\begin{align*}
\xR(t) &\triangleq \left[\begin{matrix}x_1(t) & x_2(t) & \cdots & x_{|\mathcal{R}|}(t)\end{matrix}\right]', \\
\dR(t) &\triangleq \left[\begin{matrix}d_1(t) & d_2(t) & \cdots & d_{|\mathcal{R}|}(t)\end{matrix}\right]'
\end{align*}
to be the vectors of states and subgradients of the regular nodes, respectively.  Based on Proposition~\ref{prop:LF_dynamics_regular}, the dynamics of the regular nodes under the LF dynamics can be written as
\begin{equation}\label{eq:LF_dynamics_regular_matrix}
\xR(t+1) = \AR(t)\xR(t) - \alpha_t\dR(t),
\end{equation}
where $\AR(t) \in \mathbb{R}_{\ge 0}^{|\mathcal{R}| \times |\mathcal{R}|}$ contains the weights $\bar{a}_{ij}(t)$ from \eqref{eq:LF_dynamics_regular}.


\myclearpage
\section{Convergence to Consensus}
\label{sec:consensus}

In this section, we study the convergence properties of the LF dynamics~\eqref{eq:LF_dynamics}. In particular, we provide sufficient conditions for consensus for scenarios with $ F $-local Byzantine adversaries (i.e., the most general class of adversaries that we consider), and necessary and sufficient conditions for scenarios with $ F $-total malicious adversaries. 

\subsection{A Sufficient Condition for Consensus Under $F$-local Byzantine Adversaries}

\begin{theorem}\label{thm:consensus_to_y}
Consider the network $\G = (V,\E)$, with regular nodes $\mathcal{R}$ and an $F$-local set of Byzantine nodes $\mathcal{A}$.  Suppose the network is $(2F+1)$-robust,  that the functions $f_i$, $v_i \in \mathcal{R}$, have subgradients bounded by some constant $L$, and that the regular nodes run the LF dynamics \eqref{eq:LF_dynamics} with parameter $F$.  Further suppose that $\alpha_t \rightarrow 0$ as $t \rightarrow\infty$.   Then, there exists a sequence of stochastic vectors $\mathbf{q}_t$, $t \in \mathbb{N}$, such that   
$$
\limsup_{t \rightarrow\infty}\left\| \xnet(t)- \ones y(t)\right\| =  0,
$$
where $y(t) = \mathbf{q}_t'\xnet(t)$.
\end{theorem}

\begin{IEEEproof}
Consider the LF dynamics \eqref{eq:LF_dynamics}, and their equivalent matrix representation \eqref{eq:LF_dynamics_regular_matrix}.  By Proposition~\ref{prop:LF_dynamics_regular} we know the following facts about the dynamics matrix $\AR(t)$ at each time-step $t \in \mathbb{N}$:  each diagonal element is lower bounded by $\eta$, and for each row $i \in \{1, 2, \ldots, |\mathcal{R}|\}$, at least $|\mathcal{N}_i^-| - 2F$ elements are lower-bounded by $\frac{\eta}{2}$.  Consider the graph $\G$, and remove all edges whose weights are smaller than $\frac{\eta}{2}$ in $\AR(t)$; note that this removes all edges from adversarial nodes to regular nodes (since they do not show up at all in $\AR(t)$).  For each regular node $v_i \in \mathcal{R}$, note that at most $2F$ incoming edges are removed, again since at least $|\mathcal{N}_i^-| - 2F$ elements are lower-bounded by $\frac{\eta}{2}$.  Now, from Lemma~\ref{lem:r_robust_rooted}, we see that if the graph $\G$ is $(2F+1)$-robust, the subgraph consisting of regular nodes will be rooted after removing $2F$ or fewer edges from each regular node.   Thus, $\AR(t)$ is rooted for each $t \in \mathbb{N}$, with a tree whose edge-weights are all lower-bounded by $\frac{\eta}{2}$ (and whose diagonal elements are also lower-bounded by $\frac{\eta}{2}$).    The  theorem then follows by applying the first part of  Lemma~\ref{lem:consensus_to_y}.
\end{IEEEproof}

The above proof relied on the fact that in $(2F+1)$-robust networks, the weight matrix $\bar{A}(t)$ corresponding to the regular nodes is rooted at each time-step (under the $F$-local adversary model).  This is only a sufficient condition; we now show that under the $F$-total malicious model, one can in fact give a necessary and sufficient condition on the graph topology in order to guarantee consensus, but that rootedness is no longer guaranteed at each time-step under such conditions. We will then provide an alternate proof of convergence to consensus for such graphs.

\subsection{A Necessary and Sufficient Condition for Consensus Under $F$-total Malicious Adversaries}

We start with the following example showing that when the network is not $(2F+1)$-robust, the graph induced by the filtering operation may not be rooted at each time-step.

\begin{example}\label{ex:no_spanning_tree}
Consider the graph of Figure~\ref{fig:no_span_example}(a), where all nodes are regular and use the LF dynamics \eqref{eq:LF_dynamics} with $ F=1 $. Let us assume that all nodes have identical objective functions given by $ f(x) = |x|$, and that the initial values of the nodes are as displayed inside the circles.   One can verify that this graph is only $2$-robust:  if we take each of the nodes with value $1$ to be the sets $S_1$ and $S_2$, then no node in either set has more than $2$ neighbors outside its set. Thus Theorem~\ref{thm:consensus_to_y} cannot be applied to prove consensus.  Indeed, we will show that graph induced by the LF dynamics may not be rooted at each time-step.  
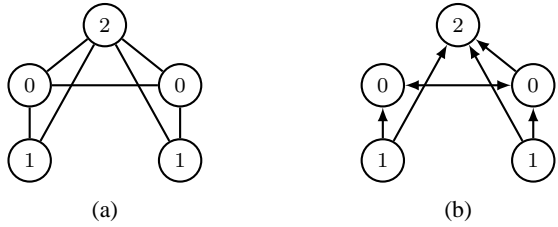
\begin{figure}[t!]
    \centering
    \begin{subfigure}[t]{0.22\textwidth}
        \centering
\begin{tikzpicture}[scale=.20, auto, node distance=1cm, thick,
   node_style/.style={circle,draw=black,fill=white!20!,font=\sffamily\Large\bfseries}]

   \node[node_style] (n1) at (0,1) {\footnotesize $2$};
   \node[node_style] (n2) at (-5, -3) {\footnotesize $0$};
   \node[node_style] (n3) at (5,-3) {\footnotesize $0$};
   \node[node_style] (n4) at (-5, -8) {\footnotesize $1$};
   \node[node_style] (n5) at (5,-8) {\footnotesize $1$};

   \foreach \from/\to in {n1/n2, n1/n3, n1/n4, n1/n5, n2/n3, n2/n4, n3/n5}
     \draw [thick] (\from) -- (\to);

    \end{tikzpicture}
     \caption{}
    \end{subfigure}%
    ~ 
    \begin{subfigure}[t]{0.28\textwidth}
        \centering
\begin{tikzpicture}[scale=.20, auto, node distance=0.5cm, thick,
   node_style/.style={circle,draw=black,fill=white!20!,font=\sffamily\Large\bfseries}]

 \node[node_style] (n1) at (0,1) {\footnotesize $2$ };
   \node[node_style] (n2) at (-5, -3) {\footnotesize $0$};
   \node[node_style] (n3) at (5,-3) {\footnotesize $0$};
   \node[node_style] (n4) at (-5, -8) {\footnotesize $1$};
   \node[node_style] (n5) at (5,-8) {\footnotesize $1$};

   \foreach \from/\to in {n3/n1,n4/n2,n4/n1,n5/n1}
     \draw [>=latex, thick, ->] (\from) -> (\to);
    
    \draw [>=latex, thick, <->] (n2) -- (n3); 
    
    \draw [>=latex, thick, <-] (n3) -- (n5); 

\end{tikzpicture}
        \caption{}
    \end{subfigure}
    \caption{\footnotesize (a) A $2$-robust network.  The values inside the circles indicate the initial values of the nodes.  (b) An arrow from node $v$ to node $w$ indicates that $w$ uses $v$'s value after applying the filtering operation.  The resulting induced graph is not rooted.}\label{fig:no_span_example}
\end{figure}
 Figure~\ref{fig:no_span_example}(b) shows the information that is used by each node after the filtering operation. For example, the node with value $ 2 $ has disregarded one of its neighbors with value $ 0 $, which is lower than its own value.  However, since the node with value $2$ does not have any neighbors with values larger than its own, it does not remove any other values.  Similarly each node with value $1$ removes the value $2$ and the value $0$, as they are the single highest and single lowest values in its neighborhood at this time-step.  The directed graph induced by the filtering operation is clearly not rooted; nevertheless, as we show later in Theorem~\ref{theorem-consensus-F-total}, the regular nodes are guaranteed to achieve consensus in this network under the dynamics \eqref{eq:LF_dynamics}, even if any single node becomes malicious.
\end{example}

This example motivates us to use a different strategy for establishing the convergence properties of the LF dynamics \eqref{eq:LF_dynamics}. More importantly, our alternate approach will allow us to show that the notion of $(r,s)$-robustness given in Definition~\ref{def:rs_robust} yields a necessary and sufficient condition for consensus in scenarios with $ F $-total malicious adversaries. In order to establish this result, we need to define the following quantities:
$$
M(t) \triangleq \max_{v_i \in \R}x_i(t), \quad m(t) \triangleq \min_{v_i \in \R}x_i(t),
$$
and 
$$
D(t) \triangleq M(t) - m(t).
$$
For each $t \in \mathbb{N}$, we set
$$
\delta_{t} \triangleq \sup_{\bar{t} \ge t}|\alpha_{\bar{t}}| L,
$$
where $L$ is the upper bound on the magnitude of the subgradients.  Clearly $|\alpha_{\bar{t}}d_i(\bar{t})| \le \delta_{t}$ for all $\bar{t} \ge t$.  For any $\gamma \in \mathbb{R}$ and $t, \bar{t} \in \mathbb{N}$ with $\bar{t} \ge t$, define the sets
\begin{align*}
\mathcal{X}_{M}(t,\bar{t},\gamma) &\triangleq \left\{v_i \in V \mid x_i(\bar{t}) > M(t) - \gamma\right\} \\
\mathcal{X}_{m}(t,\bar{t},\gamma) &\triangleq \left\{v_i \in V \mid x_i(\bar{t}) < m(t) + \gamma\right\}.
\end{align*}
A key to the proof will be the following simple fact:  at any time-step $t$, no regular node will ever use a value larger than $M(t)$ or smaller than $m(t)$ in its update equation \eqref{eq:LF_dynamics}.  This is easy to see as follows.  If the filtering operation by regular node $v_i$ removes all of the adversarial nodes in its neighborhood at time-step $t$, then clearly all remaining nodes in $\mathcal{J}_i(t)$ are regular, and thus have values in the interval $[m(t), M(t)]$.  On the other hand, if a regular node uses a value of an adversarial node, then under the $F$-local model, there must be at least one regular node in $v_i$'s neighborhood that had value larger than the value of the adversarial node, and at least one regular node in the neighborhood that had value smaller than the adversarial node's (these values could potentially be $v_i$'s own value).  Thus again, we see that all of the values used by $v_i$ at time-step $t$ are in the interval $[m(t), M(t)]$.  We are now ready to show the following result.

\begin{proposition}\label{prop:difference_decrease}
Consider the network $\G = (V,\E)$, with regular nodes $\R$ and adversarial nodes $\mathcal{A}$.  Suppose the adversarial nodes are $F$-total malicious and the network is $(F+1,F+1)$-robust.  Further suppose that the functions $f_i$, $v_i \in \R$ have subgradients bounded by some constant $L$, and that the regular nodes run the Local Filtering dynamics \eqref{eq:LF_dynamics} with parameter $F$ and with weights lower bounded by $\eta$.  Then for any $t\in\mathbb{N}$, we have
\begin{equation}
D(t + |\R|) \le \left(1 - \frac{\eta^{|\R|}}{2}\right)D(t) + 2|\R|\delta_t.
\label{eq:diff_upper_bound}
\end{equation}
\end{proposition}

\begin{IEEEproof}
 Consider any time-step $t \in \mathbb{N}$.  Define $\gamma_0 = \frac{D(t)}{2}$.  Note that the sets $\mathcal{X}_M(t,t,\gamma_0)$ and $\mathcal{X}_{m}(t,t,\gamma_0)$ are disjoint.  

By the definition of these sets, they each contain at least one regular node when $D(t) > 0$  (i.e., the nodes that have value $M(t)$ and $m(t)$, respectively).  Since the graph is $(F+1,F+1)$-robust, and since there are at most $F$ adversarial nodes, there is at least one regular node in either $\mathcal{X}_M(t,t,\gamma_0)$ or $\mathcal{X}_m(t,t,\gamma_0)$ (or both) that has $F+1$ neighbors outside its set.  Since each regular node only discards up to $F$ values that are smaller (or larger) than its own value, there will be at least one regular node that uses the value of a node from outside its set.  Suppose that there is such a regular node $v_i$ in the set $\mathcal{X}_M(t,t,\gamma_0)$. Then, the value of this node at the next time-step is upper bounded as
\begin{align*}
x_i(t+1) &\le (1-\eta)M(t) + \eta(M(t) -\gamma_0) + \delta_{t} \\
&= M(t) - \eta\gamma_0 + \delta_{t}.
\end{align*}
The above bound is obtained by noting that the smallest possible weight that a node can assign to a used value is $\eta$ (according to the description of the LF dynamics \eqref{eq:LF_dynamics}).  Note that the above expression is also an upper bound for any regular node that is not in $\mathcal{X}_M(t,t,\gamma_0)$, since such a node will use its own value in its update.

Similarly, if there is a regular node $v_j \in \mathcal{X}_m(t,t,\gamma_0)$ that uses the value of a node outside that set, then its value at the next time-step is lower-bounded as
\begin{align*}
x_j(t+1) &\ge (1-\eta)m(t)+ \eta(m(t) + \gamma_0) -\delta_{t} \\
&= m(t) + \eta\gamma_0 - \delta_{t}.
\end{align*}
Again, this is also a lower bound for the value of any regular node that is not in the set $\mathcal{X}_m(t,t,\gamma_0)$.

Now, define the quantity $\gamma_1 = \eta\gamma_0 -\delta_{t}$ and note that this is smaller than $\gamma_0$.  Thus, the sets $\mathcal{X}_M(t,t+1,\gamma_1)$ and $\mathcal{X}_m(t,t+1,\gamma_1)$ are disjoint.  Furthermore, by the bounds provided above, we see that at least one of the following must be true:
\begin{align*}
|\mathcal{X}_M(t,t+1,\gamma_1)\cap \R| &< |\mathcal{X}_M(t,t,\gamma_0)\cap \R| \\
|\mathcal{X}_m(t,t+1,\gamma_1)\cap \R| &< |\mathcal{X}_m(t,t,\gamma_0)\cap \R|.
\end{align*}
If both of the sets $\mathcal{X}_M(t,t+1,\gamma_1)\cap \R$ and $\mathcal{X}_m(t,t+1,\gamma_1)\cap \R$ are nonempty, then again by the fact that the graph is $(F+1,F+1)$-robust, there is at least one regular node in at least one of these sets that has $F+1$ neighbors outside the set.  Suppose that $v_i \in \mathcal{X}_M(t,t+1,\gamma_1)\cap \R$ is such a node.  As above, this node's value at the next time-step is upper bounded as
\begin{align*}
x_i(t+2) &\le (1-\eta)M(t+1) + \eta(M(t)-\gamma_1) +\delta_{t} \\
&\le (1-\eta)(M(t) + \delta_{t}) + \eta(M(t)-\gamma_1) +\delta_{t} \\
&= M(t) + (2-\eta)\delta_{t} - \eta\gamma_1 \\
&= M(t) + 2\delta_{t} - \eta^2\gamma_0,
\end{align*}
where the first inequality holds since the smallest possible weight that node $ v_i $ can assign to the (undiscarded) value of a neighbor outside $\mathcal{X}_M(t,t+1,\gamma_1)$ is $ \eta $, and the value of this neighbor, by construction, is at most $ M(t)-\gamma_1 $.  Again, this upper bound also holds for any regular node that is not in $\mathcal{X}_M(t,t+1,\gamma_1)\cap \R$. Similarly, if there is a node $v_j \in \mathcal{X}_m(t,t+1,\gamma_1)\cap \R$ that has $F+1$ neighbors outside that set, its next value is lower bounded as
\begin{align*}
x_j(t+2) &\ge (1-\eta)m(t+1) + \eta(m(t)+\gamma_1) - \delta_{t} \\
&\ge (1-\eta)(m(t) - \delta_{t}) + \eta(m(t)+\gamma_1) -\delta_{t} \\
&= m(t) - (2-\eta)\delta_{t} + \eta\gamma_1 \\
&= m(t) - 2\delta_{t} + \eta^2\gamma_0.
\end{align*}
This bound also holds for any regular node that is not in the set $\mathcal{X}_m(t,t+1,\gamma_1)\cap \R$.

We continue in this manner by defining $\gamma_k = \eta^{k}\gamma_0 - k\delta_{t}$.  At each time step $t+k$, if both $\mathcal{X}_M(t,t+k,\gamma_k)\cap \R$ and $\mathcal{X}_m(t,t+k,\gamma_k)\cap \R$ are nonempty, then at least one of these sets will shrink in the next time-step.  If either of the sets is empty, then it will stay empty at the next time-step, since every regular node outside that set will have its value upper bounded by $M(t) - \gamma_k$ (or lower bounded by $m(t) + \gamma_k$).  After $|\R|$ time-steps, at least one of the sets $\mathcal{X}_M(t,t+|\R|,\gamma_{|\R|})\cap \R$ or $\mathcal{X}_m(t,t+|\R|,\gamma_{|\R|})\cap \R$ is guaranteed to be empty.  Suppose the former set is empty; this means that
$$
M(t+|\R|) \le M(t) - \gamma_{|\R|}.
$$
Since $m(t + |\R|) \ge m(t)-|\R|\delta_t$, we obtain 
\begin{align*}
D(t+|\R|) &\le D(t) - \gamma_{|\R|} + |\R|\delta_t \\
&= \left(1 - \frac{\eta^{|\R|}}{2}\right)D(t) + 2|\R|\delta_t.
\end{align*}
The same expression arises if the set $\mathcal{X}_m(t,t+|\R|,\gamma_{|\R|})\cap \R$ is empty, concluding the proof.
\end{IEEEproof}

The above proposition leads to the following result for consensus of the gradient-based distributed optimization dynamics under local-filtering rules.

\begin{theorem}\label{theorem-consensus-F-total}
Consider the network $\G = (V,\E)$, with regular nodes $\R$ and an $F$-total set of malicious nodes $\mathcal{A}$.  Suppose that the functions $f_i$, $v_i \in \R$, have subgradients bounded by some constant $L$, and that the regular nodes run the Local Filtering dynamics \eqref{eq:LF_dynamics} with parameter $F$ and weights lower bounded by $\eta$.  Suppose further that the step-sizes satisfy $\alpha_t \rightarrow 0$.  Then the regular nodes are guaranteed to reach consensus despite the actions of the adversaries, initial values, and local functions if and only if the graph is $(F+1,F+1$)-robust.
\end{theorem}

\begin{IEEEproof}
The proof of sufficiency follows immediately from Proposition~\ref{prop:difference_decrease} by noting that when $\alpha_t \rightarrow 0$, then $\delta_t \rightarrow 0$ as $t \rightarrow 0$.  By input-to-state stability, we have $D(t) \rightarrow 0$ as $t \rightarrow 0$, proving consensus.

For necessity, suppose that the network is not $(F+1,F+1)$-robust.  Then there exist two disjoint nonempty sets $S_1, S_2 \subset V$ such that (i) there is at least one node in $S_1$ that has at most $F$ neighbors outside $S_1$, (ii) there is at least one node in $S_2$ that has at most $F$ neighbors outside $S_2$, and (iii) there are at most $F$ nodes in $S_1 \cup S_2$ that have $F+1$ or more neighbors outside their respective sets.  Choose the nodes in $S_1 \cup S_2$ that each have $F+1$ or more neighbors outside their respective sets to be the adversarial set $\mathcal{A}$; clearly $\mathcal{A}$ is an $F$-total set.  Now, assign all of the nodes in set $S_1$ to have function $f_1$, and assign all of the nodes in set $S_2$ to have function $f_2$, where the minimizer of $f_2$ is strictly larger than the minimizer of $f_1$.  Now let all of the nodes in set $V \setminus \{S_1 \cup S_2\}$ have function $f_3$, selected to have gradient equal to zero in the entire interval bracketed by the minimizers of $f_1$ and $f_2$.   Let all nodes in $S_1$ and $S_2$ (including the adversarial nodes) be initialized at their local minimizers, and let all nodes in $V \setminus \{S_1 \cup S_2\}$ be initialized at a value strictly between the minimizers of $f_1$ and $f_2$.  Furthermore, let the malicious nodes never change their values.  In this case, all regular nodes in $S_1$ will discard all of their neighbors' values from outside $S_1$ (since they each have at most $F$ neighbors outside $S_1$), and similarly all regular nodes in $S_2$ will discard all of their neighbors' values from outside $S_2$.  As the values of nodes in $V \setminus \{S_1 \cup S_2\}$ will always remain strictly between the minimizers of $f_1$ and $f_2$, the regular nodes in $S_1$ and $S_2$ will never deviate from their initial values, and thus consensus will not be reached.  
\end{IEEEproof}

The above result shows that the network considered in Example~\ref{ex:no_spanning_tree} is guaranteed to facilitate consensus among the regular nodes despite the presence of any single malicious node (since the network is $(2,2)$-robust), even though the graph induced by the filtering operation is not rooted at each time-step.

\begin{remark}
As illustrated by Theorems~\ref{thm:consensus_to_y} and \ref{theorem-consensus-F-total}, the properties of $r$-robustness and $(r,s)$-robustness play a key role in consensus-based optimization dynamics of the form \eqref{eq:LF_dynamics}.  While these properties are stronger than other graph properties such as $r$-minimum degree and $r$-connectivity, all of these properties occur simultaneously in various commonly studied models for large-scale networks \cite{HZ-EF-SS:15}.  There are also various simple techniques to construct $r$-robust networks for any given $r \in \mathbb{N}$, as discussed in \cite{HJL-HZ-XK-SS:13}. 
\end{remark}


\myclearpage
\section{A Safety Condition:  Convergence to the Convex Hull of the Local Minimizers}
\label{sec:safety}

In the previous section, we provided graph properties that guaranteed consensus for the regular nodes under the LF dynamics \eqref{eq:LF_dynamics} (under the condition that the step-sizes asymptotically go to zero).  In this section, we provide a {\it safety guarantee}  on these dynamics under additional conditions on the step-sizes, as detailed in the following theorem.  

\begin{theorem} 
Suppose that one of the following conditions holds:
\begin{enumerate}
\item The adversarial nodes are $F$-total malicious and the network is $(F+1,F+1)$-robust; or
\item The adversarial nodes are $F$-local Byzantine and the network is $(2F+1)$-robust.
\end{enumerate}
Suppose that all regular nodes follow the LF dynamics \eqref{eq:LF_dynamics} with parameter $F$.  For each node $v_i \in \mathcal{R}$, let the local function $f_i$ be convex, continuous and have subgradients bounded by $L$.   Let the set of minimizers of $f_i$ be denoted by the closed convex set  $\mathcal{M}_i \subseteq \mathbb{R}$.   Define $\overline{M} = \max_{v_i \in \mathcal{R}} \max  \{x \mid x \in \mathcal{M}_i\}$ and $\underline{M} = \min_{v_i \in \mathcal{R}} \min  \{x \mid x \in \mathcal{M}_i\}$.  If the step-sizes satisfy $\sum \alpha_t = \infty$ and $\alpha_t \rightarrow 0$, then $\limsup_{t \rightarrow \infty}x_i(t) \le \overline{M}$ and $\liminf_{t \rightarrow \infty}x_i(t) \ge \underline{M}$ for all $v_i \in \mathcal{R}$, regardless of the actions of the adversarial nodes and the initial values.  
\label{thm:safety}
\end{theorem}

\begin{IEEEproof}
Let $M(t)$ and $m(t)$ be the maximum and minimum values of the regular nodes at time-step $t$, respectively.  Under the conditions of the theorem, Theorems~\ref{thm:consensus_to_y} and \ref{theorem-consensus-F-total} indicate that $M(t)-m(t) \rightarrow 0$.  Now consider the local filtering dynamics \eqref{eq:LF_dynamics}.  Since no regular node ever adopts a neighbor's value larger than $M(t)$ in its update, we have
\begin{align*}
x_i(t+1)&= a_{ii}(t)x_i(t) + \sum_{v_j \in \mathcal{J}_i(t)} a_{ij}(t) x_j(t) - \alpha_td_i(t), \\
 &\le a_{ii}(t)M(t) + \sum_{v_j \in \mathcal{J}_i(t)} a_{ij}(t) M(t) - \alpha_td_i(t)\\
 &= M(t) - \alpha_td_i(t),
\end{align*} 
for each regular node $v_i \in \mathcal{R}$.  In particular, we have 
\begin{equation}
M(t+1) \le M(t)-\alpha_t \min_{v_i \in \mathcal{R}}d_i(t).
\label{eq:max_value_bound_1}
\end{equation}
Iterating, we obtain for any $T \in \mathbb{Z}_{\ge 1}$, 
\begin{equation}
M(t+T) \le M(t) - \sum_{j = t}^{t+T-1}\alpha_j\min_{v_i \in \mathcal{R}}d_i(j).
\label{eq:max_value_bound}
\end{equation}
    
Now suppose by way of contradiction that $\limsup_{t \rightarrow \infty}M(t) = \overline{M} + \delta$ for some $\delta > 0$.  Let $t_0$ be such that the following three conditions are satisfied: 
\begin{enumerate}
\item $\overline{M} + \frac{\delta}{2} \le M(t_0) \le \overline{M}+2\delta$,
\item $M(t)-m(t) \le \frac{\delta}{4}$ for all $t \ge t_0$, and
\item $\alpha_tL \le \frac{\delta}{4}$ for all $t \ge t_0$.
\end{enumerate}
Such a $t_0$ is guaranteed to exist by the convergence of $M(t)-m(t)$ to zero and the definition of $\delta$.  Define
$$
G= \min_{v_i \in \mathcal{R}}\left.\frac{df_i}{dx}\right|_{\overline{M}+\frac{\delta}{4}}.
$$
If $f_i$ is not differentiable at $\overline{M}+\frac{\delta}{4}$, we consider the infimum of its subgradients at that point (note that all such subgradients will be positive and bounded away from zero). Thus, we have $d_i(t) \ge G > 0$ whenever $m(t) \ge \overline{M}+\frac{\delta}{4}$.  By the definition of $t_0$ and using \eqref{eq:max_value_bound}, we have
\begin{align*}
M(t_0+T) &\le  M(t_0) - G\sum_{j = t_0}^{t_0+T-1}\alpha_j \\
&\le \overline{M} + 2\delta- G\sum_{j = t_0}^{t_0+T-1}\alpha_j, 
\end{align*}
for any $T$ such that $M(t) \ge \overline{M}+\frac{\delta}{2}$ for all $t \in [t_0, t_0+T]$.  Thus, using the fact that $\sum_{j = t_0}^{t_0+T-1}\alpha_j$ is unbounded in $T$, we see that $M(t_0+T) \le \overline{M}+\frac{\delta}{2}$ for sufficiently large $T$.  Let $t_1$ be that point in time.  

Now we show that $M(t)$ will never exceed $\overline{M} + \frac{3\delta}{4}$ after time $t_1$.  Specifically, if $M(t) \le \overline{M} + \frac{\delta}{2}$ at some time $t \ge t_1$, then by \eqref{eq:max_value_bound_1}, we have
$$
M(t+1) \le M(t) + \alpha_tL \le \overline{M} + \frac{\delta}{2} + \frac{\delta}{4}=  \overline{M} + \frac{3\delta}{4}.
$$
Similarly, if $M(t) \ge \overline{M} + \frac{\delta}{2}$ at some time $t \ge t_1$, then by \eqref{eq:max_value_bound_1}, we have $M(t+1) \le M(t) - \alpha_tG$, and thus $M(t)$ will monotonically decrease until it is below $\overline{M} + \frac{\delta}{2}$.  Thus, $M(t)$ will eventually be upper bounded by $\overline{M} + \frac{3\delta}{4}$, contradicting the definition of $\delta$.  Thus, $\limsup_{t \rightarrow\infty}M(t) \le \overline{M}$.  An identical argument holds for the lower bound.
\end{IEEEproof}

\subsection{Lack of Convergence to a Constant Value Under Adversarial Behavior}
As shown in the previous result, the LF dynamics guarantee consensus within the convex hull of the local minimizers and prevent the adversarial nodes from driving the states of regular nodes to arbitrarily large values under appropriate conditions on the network topology.  However, a single malicious node can still prevent the regular nodes from converging to a \emph{constant} value under certain classes of step-sizes.  This is illustrated in the following example.    

\begin{figure}
\centering
\includegraphics[width=8cm]{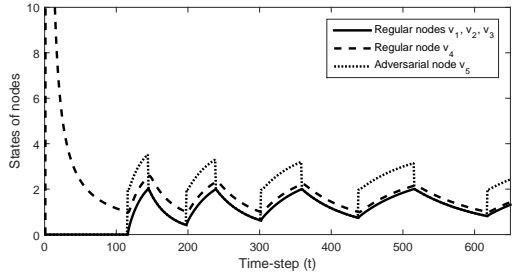}
\caption{An illustration of lack of convergence to a constant value under adversarial behavior.}
\label{fig:no_converge}
\end{figure}

\begin{example}
Consider a complete graph $\G$ with five nodes $\{v_1, v_2, v_3, v_4, v_5\}$.  Suppose $v_1, v_2$ and $v_3$ all have local functions $f_a(x) = x^2$, and $v_4$ has local function $f_b(x) = (x-9)^2$ (with the magnitude of their gradients capped at $L$, for some sufficiently large $L$).  Suppose node $v_5$ is malicious.  

Let all regular nodes start at their local minimizers and run the dynamics \eqref{eq:LF_dynamics} with step-sizes satisfying $\sum_t\alpha_t = \infty$ and $\sum_t \alpha_t^2 < \infty$.  Let the malicious node behave as follows (illustrated in Figure~\ref{fig:no_converge}).  It starts by keeping its value the same as the regular nodes $v_1, v_2$ and $v_3$.  In this case, those regular nodes all discard node $v_4$'s value as being too extreme, and thus all regular nodes  converge towards the minimizer of $f_a$, namely $0$.  When node $v_4$'s value is sufficiently close to $0$, the malicious node switches its value to be larger than $v_4$'s value (as shown just after time-step $100$ in Figure~\ref{fig:no_converge}).  At this point, all regular nodes discard $v_5$'s value as being too extreme and incorporate node $v_4$'s values in their updates.  This causes all regular nodes to start converging towards the minimizer of some convex combination of $f_a$ and $f_b$.  When all regular nodes are sufficiently close to this minimizer, the malicious node again switches its value to be the same as that of $v_1, v_2$ and $v_3$.  These three nodes then start ignoring $v_4$'s value, which causes all regular nodes to start converging towards the minimizer of $f_a$.  By repeating this behavior ad infinitum, the malicious node causes the regular nodes to forever oscillate between two different values (although they reach consensus and remain within the convex hull of the local minimizers of the regular nodes), as shown in Figure~\ref{fig:no_converge}.  
\end{example}

A formal proof of the behavior exhibited by the above example is straightforward but tedious, and thus we omit it in the interest of space.

\myclearpage
\section{Factors that Affect the Performance of Resilient Distributed Optimization Algorithms}
\label{sec:performance}

The proof of Theorem~\ref{thm:fund} indicates that the nature of the individual optimization functions (together with the network topology) will play a role in determining the performance that is achievable under adversarial behavior.  For example, suppose that all individual objective functions are drawn from a certain class of functions $\funcset$.   In the trivial case where all functions in $\funcset$ have the same minimizer, each node can calculate the globally optimal value simply by calculating the minimizer of its own function, and thus resilience to any number of adversarial nodes is guaranteed.  On the other hand, when the class of functions $\funcset$ is sufficiently rich so that the function held by each node contributes to the global minimizer, then the number and location of adversarial nodes will play a larger role in determining the achievable performance.   One such bound on performance is provided by the following result.  

\begin{proposition}
Consider a network $\G = (V,\E)$ with $n$ nodes and let $F \in \mathbb{N}$.  Let $\mathcal{T}\subset V$ be a maximum $F$-local set.  Let $\funcset$ be the set from which the local objective functions are drawn, and suppose that $ f_a,f_b \in \funcset $, where $f_a(x)=(x-a)^2 $ and $f_b(x)=(x-b)^2 $, for $a,b\in \real$. \footnote{Both functions can be modified to have their gradients capped at sufficiently large values, so as to not affect the minimizer of any convex combination of the functions.} Let $\Gamma$ be any distributed optimization algorithm that guarantees that all regular nodes reach consensus on a value in the convex hull of the minimizers of the regular nodes' objective functions.   Let $x^*$ be the true minimizer of the average of the functions held by all regular nodes, and let $\bar{x}$ be the value computed by the regular nodes under $\Gamma$.  Then, under the $F$-local adversary model, there is an allocation of functions to nodes such that $|\bar{x} - x^*| = \frac{|\mathcal{T}|}{n}|(b-a)|$ and $f(\bar{x})-f(x^*) = \frac{|\mathcal{T}|^2}{n^2}(b-a)^2$, where $f(x)$ is the value of the average of the functions held by the regular nodes evaluated at $ x $.
\label{prop:F_local_lower_bound}
\end{proposition}

\begin{IEEEproof}
We consider two scenarios.  In the first scenario, let each node in $V \setminus \mathcal{T}$ have the local function $f_a$, and let each node in $\mathcal{T}$ have the local function $f_b$.  Let all nodes be regular.  The minimizer of the average of all functions is given by $x^* = a+\frac{|\mathcal{T}|(b-a)}{n}$, with $f(x^*) = \left(1-\frac{|\mathcal{T}|}{n}\right)\frac{|\mathcal{T}|}{n}(b-a)^2$.  

In the second scenario, the nodes in set $\mathcal{T}$ are also assigned the function $f_a$, but are adversarial and execute the algorithm by pretending their local functions are $f_b$.  Since $\Gamma$ guarantees that all regular nodes reach consensus in the convex hull of the minimizers of the regular nodes' functions, all regular nodes must obtain the value $\bar{x} = a$ after executing algorithm $\Gamma$.   

Since the two scenarios are indistinguishable from the perspective of $\Gamma$, the algorithm must also cause all regular nodes to calculate $\bar{x} = a$ under the first scenario.  Thus, the difference of the value output by $\Gamma$ and the true minimizer of the regular nodes' functions is $|\bar{x} - x^*| = \frac{|\mathcal{T}|}{n}|(b-a)|$, and the difference in achieved costs is $f(\bar{x})-f(x^*) = \frac{|\mathcal{T}|^2}{n^2}(b-a)^2$.
\end{IEEEproof}

\begin{figure}[bt]
\begin{center}
\begin{tikzpicture}  [scale=.16, inner sep=1pt, minimum size=20pt, auto=center, node/.style={circle, draw=black, thick}]
  \node [circle, draw, fill=white](v1) at (0,0)  {\footnotesize $w_1$};
  \node [circle, draw, fill=white](v2) at (5,0)  {\footnotesize $w_2$};
  \node [circle, draw, fill=white](v3) at (10,0)  {\footnotesize $w_3$};
  \node [circle, draw, fill=white](v4) at (15,0)  {\footnotesize $w_4$};
  \node [circle, draw, fill=white](v5) at (20,0)  {\footnotesize $w_5$};
  \node [circle, draw, fill=white](v6) at (25,0)  {\footnotesize $w_6$};
  \node at (30,0) { $\cdots$};
  \node [circle, draw, fill=white](v7) at (35,0)  {\tiny $w_{N-2}$};
  \node [circle, draw, fill=white](v8) at (40,0)  {\tiny $w_{N-1}$};
  \node [circle, draw, fill=white](v9) at (45,0)  {\footnotesize $w_{N}$};
  
  \node [circle, draw, fill=white](w1) at (5,8)  {\footnotesize $u_1$};
  \node [circle, draw, fill=white](w2) at (20,8)  {\footnotesize $u_2$};
  \node at (30,8) {$\cdots$};
  \node [circle, draw, fill=white](w3) at (40,8)  {\footnotesize $u_K$};

  \foreach \from/\to in {w1/v1,w1/v2,w1/v3,w2/v4,w2/v5,w2/v6,w3/v7,w3/v8,w3/v9}
     \draw [thick] (\from) -- (\to);

       \foreach \from/\to in {v1/v2,v2/v3,v3/v4,v4/v5,v5/v6,v7/v8,v8/v9}
     \draw [thick] (\from) -- (\to);
     
     \draw [thick] (v6) -- (28,0);
     \draw [thick] (v7) -- (32,0);
          

  \draw[rounded corners, dashed] (-3, -3) rectangle (48, 3) {};
  \draw[rounded corners, dashed] (2, 5) rectangle (43, 11) {};
  \node at (51,0) {$W$};
  \node at (48,8) {$U$};

\end{tikzpicture}    
\end{center}
\caption{Graph $\G$ constructed on node sets $U\cup W$.  All nodes in set $W$ are connected to each other (the edges are not shown in the interest of clarity).  Each node in set $U$ connects to three unique vertices in set $W$.  This graph is $3$-robust.}
\label{fig:graph_performance}
\end{figure}
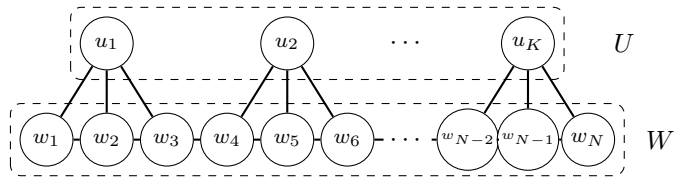

\begin{example}
Consider the network shown in Figure~\ref{fig:graph_performance}, where $K \ge 2$ is some positive integer, and $N = 3K$.  We define the vertex sets $W = \{w_1, w_2, \ldots, w_N\}$ and  $U = \{u_1, u_2, \ldots, u_K\}$.  One can verify that this network is $3$-robust and that $U$ is a maximum $1$-local set.  Suppose each node in $U$ is assigned the function $f_b(x) = (x-b)^2$, and each node in $W$ is assigned the function $f_a(x) = x^2$.  By Proposition~\ref{prop:F_local_lower_bound}, any algorithm that guarantees to output a value in the convex hull of the regular nodes' minimizers must produce $\bar{x} = 0$ as a solution.  In this case, we have $|\bar{x}- x^*| = \frac{b}{4}$ and $f(\bar{x})-f(x^*) = \frac{b^2}{16}$, where $x^* = \frac{b}{4}$ is the global minimizer.  
\end{example}

Given the fact that the performance of resilient distributed optimization algorithms heavily depends on the size of maximum $F$-local sets in the network (under the $F$-local adversary model), it is natural to ask how easy it is to find such maximum sets.  To answer this, we first define the problem formally and then characterize its complexity.

\begin{definition}
Let $r, k$ be positive integers.  The {\it $r$-Local Set Problem} is to determine whether a given graph has an $r$-local set of size at least $k$.
\end{definition}

\begin{theorem}
  The $r$-Local Set Problem is NP-complete.
\label{thm:r_local_NP_complete}
\end{theorem}

The proof of the above theorem is given in Appendix~\ref{sec:f_local_np_hardness}.

Although finding maximum $F$-local sets in graphs is difficult in general (unless $P = NP$), one can characterize the size of such sets in certain specific classes of graphs.  For instance, the maximum $F$-local set in complete graphs has size exactly $F$.  Similarly, consider Erd{\"o}s-R\'{e}nyi random graphs where each edge between each pair of nodes is added independently with a certain probability $p(n)$ (which could depend on the number of nodes in the graph).  It was shown in \cite{HZ-EF-SS:15, SJ-TL-TT-TV:12} that if the edge probability satisfies
\[
p(n)=\frac{\ln(n)+F\ln\ln(n)+g(n)}{n},
\] 
where $ g(n) \rightarrow \infty $ as $ n \rightarrow \infty $, the size of the largest $ F $-local set is in $ O(n\gamma(n) ) $ with high probability, where $\gamma(n)$ is any function satisfying $\ln\ln(n) = o(\gamma(n) \ln n)$.  For instance, $\gamma(n) = \frac{(\ln\ln(n))^{1+\epsilon}}{\ln(n)}$ satisfies this for any $\epsilon > 0$.  Thus, with high probability, the fraction of nodes that are in the maximum $F$-local set goes to zero as $n \rightarrow \infty$ in Erd{\"o}s-R\'{e}nyi random graphs for the above regime of edge probabilities.  This means that the limitation identified in Proposition~\ref{prop:F_local_lower_bound} will not play a major role in such graphs.  An interesting avenue for further research is to identify whether there are other graph theoretic obstructions to the performance of resilient distributed optimization algorithms (including the LF dynamics we have presented in this paper).

\myclearpage
\section{Directions for Future Research}
\label{sec:conclusion}
In this paper, we proposed a consensus-based distributed optimization algorithm that is resilient to adversarial behavior under certain conditions on the network topology, in the sense that the regular nodes will always asymptotically converge to the convex hull of the minimizers of the regular nodes' functions, despite the actions of any $F$-local set of adversaries.  We also identified topological properties (in the form of maximum $F$-local sets) that affect the performance of the algorithm.  There are many interesting directions for future research, including a more explicit characterization of the distance-to-optimality of such algorithms (with corresponding conditions on the network topology), along with a characterization of classes of functions that lead to near-optimal solutions.

 \myclearpage
\bibliographystyle{IEEEtran}
\bibliography{alias,refs}

\appendices

\section{Proof of Theorem~\ref{thm:r_local_NP_complete}:  Complexity of Finding Maximum $r$-Local Sets}
\label{sec:f_local_np_hardness}

\begin{IEEEproof}
We will provide a reduction from the NP-complete Set Packing problem:  given a collection of elements $U = \{u_1, u_2, \ldots, u_n\}$, a set of subsets $S = \{S_1, \ldots, S_m\}$ of $U$, and a positive integer $k$, do there exist $k$ subsets in $S$ that are mutually disjoint?  Specifically, we will show that given any instance of the Set Packing problem, one can construct a graph $\G = (V, \E)$ in such a way that $\G$ contains a $1$-local set of size at least $k$ if and only if the answer to the given instance of the Set Packing problem is ``yes.''  We assume throughout that $k \ge 2$, as the answer to the Set Packing problem for $k = 1$ is always ``yes.''

Construct the graph $\G$ as follows.  Define the vertex set $V$ to consist of $n + m$ vertices 
$$
V = \{u_1, u_2, \ldots, u_n, s_1, s_2, \ldots, s_m\}, 
$$
where each vertex $u_i$ corresponds to an element of the set $U$, and each vertex $s_i$ corresponds to the subset $S_i \in S$.  

Next, place an edge between each pair of vertices $u_i, u_j$, $j \ne i$.  This creates a complete graph on the vertex set $\{u_1, \ldots, u_n\}$.  For each vertex $s_i$, $1 \le i \le m$, add an edge between $s_i$ and vertex $u_j$ if $u_j \in S_i$ in the given instance of the Set Packing problem.    This completes the construction of the graph $\G$.  

Suppose that the answer to the Set Packing instance is ``yes.'' Then there exists a collection of at least $k$ subsets such that no two of the subsets share an element.  Let $\mathcal{P} = \{S_{i_1}, S_{i_2}, \ldots, S_{i_{k'}}\}$ be the corresponding collection, where $k' \ge k$.  Let $\mathcal{P}_v = \{s_{i_1}, s_{i_2}, \ldots, s_{i_{k'}}\} \subset V$ be the corresponding vertices in graph $\G$.  Then it is easy to verify that $\mathcal{P}_v$ forms a $1$-local set of size $k' \ge k$; none of the vertices $\{u_1, u_2, \ldots, u_n\}$ have more than one neighbor in $\mathcal{P}_v$ (by the definition of the edges and the fact that $\mathcal{P}_v$ corresponds to a packing), and none of the vertices $s_i$ share any edges with nodes in the set $\mathcal{P}_v$.  Thus, if the answer to the Set Packing instance is ``yes'', the answer to the constructed instance of the $1$-local Set Problem is ``yes.''

We now show the converse.  Suppose the answer to the constructed instance of the $1$-local Set Problem is ``yes,'' i.e., there exists a $1$-local set $\mathcal{P}_v \subset V$ of vertices, with cardinality $k' \ge k \ge 2$.  We first claim that $\mathcal{P}_v$ cannot contain any vertices from the set $\{u_1, u_2, \ldots, u_n\}$.  To see this, note that $\mathcal{P}_v$ cannot contain all of the vertices $\{u_1, u_2, \ldots, u_n\}$, for if it did, any vertex $s_i$ that is not in $\mathcal{P}_v$ would contain at least two neighbors in $\mathcal{P}_v$ contradicting the fact that it is a $1$-local set.  Next, note that $\mathcal{P}_v$ cannot contain more than one node from $\{u_1, u_2, \ldots, u_n\}$, for if it did, any node $u_j$ that is not in $\mathcal{P}_v$ would have more than one neighbor in $\mathcal{P}_v$, again contradicting the fact that it is a $1$-local set.  Thus suppose $\mathcal{P}_v$ contains a single vertex from $\mathcal\{u_1, \ldots, u_n\}$, and take this vertex to be $u_i$. Then each vertex $u_j$ ($j \ne i$) already has a neighbor in $\mathcal{P}_v$, and thus none of the vertices $s_i$, $1 \le i \le m$ can be in $\mathcal{P}_v$.  Thus $\mathcal{P}_v$ is of size $1$, contradicting the fact that it is a $1$-local set of size at least $2$.    

Thus, $\mathcal{P}_v$ can contain only vertices from the set $\{s_1, s_2, \ldots, s_m\}$.  It is now easy to see that the subsets from the Set Packing problem corresponding to those vertices form a packing of size at least $k$, and thus the answer to the Set Packing problem is ``yes.''

The above reduction shows that the $r$-local Set Problem is NP-hard.  Since this problem has a certificate for ``yes'' instances that can be verified in polynomial time (i.e., the actual $r$-local set of size at least $k$), the $r$-local Set Problem is in NP, and thus is NP-complete.
\end{IEEEproof}

\end{document}